\documentclass[preprint]{revtex4}

\begin{document}
\title{\bf Energy Functional dependence of exchange coupling and magnetic properties of Fe/Nb multilayers }

\author{ {Nitya Nath Shukla and R. Prasad }
\\
\em
{Department of Physics,}
\\
{Indian Institute of Technology, Kanpur 208016, India}
}

\begin{abstract}
We present an $\it{ab}$ $\it{initio}$ calculation of the exchange coupling for 
Fe/Nb multilayers using the self-consistent full-potential linearized 
augmented-plane wave (FLAPW) method. 
The exchange correlation potential has been treated in the local spin density approximation (LSDA) as well as generalized 
gradient approximation (GGA). 
We find that for the LSDA as well as the GGA the exchange coupling oscillates with
 a period of 6.0 \AA\ of Nb spacer thickness which is
close to the experimental value in the preasymptotic region. 
This is also close to the earlier
calculated period ( i.e. 4.5 \AA\ ) by augmented spherical wave (ASW) method. 
The LSDA shows antiferromagnetic coupling for 2 and 5 Nb monolayers (ML) 
but the GGA shows the ferromagnetic coupling for all Nb spacer layers.
The period of oscillation is found to be in good agreement with the period calculated using  
the Ruderman-Kittel-Kasuya-Yosida (RKKY) and quantum well (QW) models. The 
magnetic moment of Fe is found to be higher in the GGA than the LSDA. Fe magnetic
moment also shows strong oscillations as a function of the spacer layer 
thickness, in agreement with the experimental results. We find that the GGA results
show better agreement with the experiment than the LSDA results.
\end{abstract}
\maketitle

\section{ Introduction}
During the last decade, magnetic multilayers have received a lot of attention due 
to their interesting properties like oscillating interlayer exchange coupling (IEC)
 between antiferromagnetic (AF) and ferromagnetic (FM) ordered layers\cite{ kn:Grunberg,kn:Parkin,kn:Unguris} and giant magnetoresistance
(GMR) apart from their industrial applications. 
Such oscillations in interlayer exchange coupling and the saturation 
magnetoresistance were reported by Parkin $\it{et}$ $\it{al.}$ \cite{kn:Parkin} 
with a period 15-20 \AA\ in Fe/Cr, Co/Cr and Co/Ru multilayers. Purcell 
$\it{et}$ $\it{al.}$ \cite{kn:Purcell} showed that the coupling oscillates with a smaller
 period of 3.0-4.0 \AA\ in Fe/Cr/Fe sandwich structure because of the high
degree of perfection of Fe whisker substrate and the sharpness of Fe/Cr
 interfaces. The first theoretical
 explanation was provided by Bruno $\it{et}$ $\it{al.}$ \cite{kn:Bruno}, 
who explained the oscillations as a function of spacer 
layer thickness due to the Ruderman-Kittel-Kasuya-Yosida (RKKY) interaction 
between the magnetic layers. They expressed the period of oscillations in 
terms of nesting vectors of the spacer's Fermi surface.
However, the coupling strength J 
was described by an adjustable parameter in their work. There are several 
other explanations of this phenonmenon based on electronic Fabry-perot-like interference effects 
\cite{kn:Bruno1} or quantum-well (QW) theories\cite{kn:ortega}. 
In order to understand this phenonmenon, first-principles techniques have also 
been employed which are more transparent and parameter free but require 
considerable computational effort.
The first-principles calculations
 of the exchange coupling J have been reported for various systems such as 
Fe/Cr, Fe/Cu, Fe/Mo, Co/Cu \cite{kn:Lang, kn:Mirbt, kn:Mirbt1, kn:Niklasson}. 
Recently, the combined theoretical-experimental work on Fe/Au supercells, 
with only 2 monolayers of Fe  
\cite{kn:Yoshihara}, showed that the exchange coupling J exhibits oscillatory 
behaviour. However, this was found to be ferromagnetic for all layers. 
Some results have also been  
reported for the oscillatory behaviour of magnetic moments as a function 
of spacer layer thickness \cite{kn:Mirbt}.

Fe/Nb is an interesting system as it provides a way of exploring the 
coexistence of ferromagnetism and superconductivity. This is one of the main 
interests of experimental 
studies of Fe/Nb/Fe multilayers\cite{kn:Mattson, kn:Klose, kn:Muhge, kn:Rehm}.
M$\ddot{u}$hge $\it et. al.$ \cite{kn:Muhge} studied the magnetism and superconducting 
properties of Fe/Nb trilayer system. They have found a strong 
dependence of the superconducting transition temperature $T_c$ on Fe layer 
thickness. They also studied the magnetic properties using the magneto-optical 
Kerr effect, ferromagnetic resonance and magnetization measurements by SQUID 
magnetometer. 

Although a large number of studies exist for 
Fe/Cr, Fe/Au, Co/Cr $\it etc.$, 
there are only a few theoretical studies for Fe/Nb system. 
The difficulty one faces with Fe/Nb is that the lattice parameter of Fe and Nb
differ by $13\%$ and thus it presents an additional complication for theoretical 
as well as experimental studies.
The Fe/Nb multilayers show strong
 interlayer exchange coupling \cite{kn:Mattson} which  changes in continuous 
and reversible way by introducing hydrogen into the sample \cite{kn:Klose}.
This system has not been explored much theoretically and therefore needs a detailed study.
The experimental results for Fe/Nb multilayers \cite{kn:Mattson, kn:Muhge, kn:Rehm} show 
oscillatory interlayer exchange coupling as a function of Nb spacer layer 
thickness within a period of 9.0 \AA\ to 12.8 \AA\ . Sticht $\it{et}$ $\it{al.}$ \cite{kn:Sticht}
 showed the oscillating exchange coupling in Fe/Nb multilayers as a function of Nb spacer layer thickness using the augmented spherical wave (ASW) method
\cite{kn:Williams}. 

To study the exchange coupling in Fe/Nb multilayers, we have used a self-consistent full-potential linearized augmented plane-wave (FLAPW) method \cite{kn:Blaha} that  
treats crystal potential accurately and makes no shape approximation 
for the crystal potential.
In this paper we report the energy functional dependence of exchange coupling 
and magnetic properties of Fe/Nb multilayers. Some preliminary results of our
work have been reported earlier \cite{kn:nitya}.
Our results show that the interlayer exchange coupling for both the local 
spin density approximation (LSDA) and the generalized gradient approximation 
(GGA) oscillates with a period of 6.0 \AA\ which is close to  
the ASW result of 4.5 \AA\ \cite{kn:Sticht}. Our calculated period is close to 
the experimental result in the preasymptotic region. We find that the GGA favors FM coupling
for all Nb monolayers while the LSDA gives FM as well as AF coupling.
We have also studied the change in magnetic 
moment of Fe and Nb as a function of Nb spacer layer thickness within 
the LSDA and the GGA.  
We find that Nb develops induced magnetic 
moments which oscillate as a function of Nb spacer layer. 
The magnetic moment on Fe gets reduced appreciably from its bulk value and 
depends on Nb spacer layer thickness. 
The Fe magnetic moment shows strong oscillations as a function of Nb thickness
in agreement with the experimental results
\cite{kn:Mattson, kn:Muhge}.

The organization of the paper is as follows. 
In Sec. II we provide the details of our calculation. In Sec. III we present 
our results and discuss them. Finally, we give our conclusions in Sec. IV.

\section{Computational details}
\subsection{Method}
All calculations reported in this paper have been carried out using the self-consistent
 full-potential linearized augmented plane-wave (FLAPW) method 
\cite{kn:Blaha} in a scalar
relativistic version without spin-orbit coupling. We have used the LSDA as well 
as GGA for our calculations. The muffin-tin radii are 1.33 \AA\ for Fe and Nb.
 The maximum value in the radial sphere expansion is $l_{max} = 10$, and the 
largest $l$ value for the non-spherical part of the Hamiltonian matrix is 
$l_{max,ns} = 4$. The cutoff parameters are $R_{mt}K_{max} = 8$ for the plane 
wave and $R_{mt}G_{max} = 14$ for the charge density. 
The number of plane waves ranges from 1400 to 5475, depending
 on the number of Nb layers.  An improved tetrahedron method has been used for 
the Brillouin-zone integration \cite{kn:blochl}. 

\subsection{Supercell geometry}
To perform the calculations we have 
constructed tetragonal supercells consisting of bcc Fe and Nb monolayers 
and in each supercell the number of Nb monolayers (ML) range from 1ML to 7ML. 
The schematic geometry of $FeNb_{4}$ multilayers is shown in Fig. 1. The 
lattice constants of bulk Fe and Nb are very different with 
$13\%$ lattice mismatch. 
To reduce the lattice mismatch we started with a multilayer whose lattice 
constant is average of bulk Fe and Nb and then relaxed the lattice parameter. 
The relaxed parameter was found to be 3.067 \AA\ . 
All calculations are performed with this relaxed lattice constant. 
The lattice parameter along (001) 
direction changes from 3.067 \AA\ to 12.269 \AA\ in supercells. The atomic 
positions have not been relaxed. 

\subsection{Interlayer exchange coupling and $\bf k$-point convergence}

The total energy is calculated for ferro 
and antiferromagnetic ordering of Fe atoms for each Nb spacer layer thickness. 
To obtain reliable energy differences between ferromagnetic and 
antiferromagetic ordering, the unit cell of AF structure is also used for FM 
structure.
The exchange coupling, J, is calculated by taking the energy difference
\begin{equation}
J(d) = E_{tot}^{\uparrow\downarrow}(d) - E_{tot}^{\uparrow\uparrow}(d)
\end{equation}
where $d$ is the thickness of the spacer layer and $E_{tot}^{\uparrow\downarrow}(d)$
 and $E_{tot}^{\uparrow\uparrow}(d)$ are the total energies of the system in 
antiferromagnetic and ferromagnetic arrangements.
For the calculation of the IEC, a sufficiently large number of
$\textbf{k}$ points are needed. 
The convergence of the IEC as a function of $\textbf{k}$ points 
in the irreducible Brillouin zone (IBZ) is shown in Fig. 2 for the LSDA and GGA 
potentials. We find that the convergence 
is reached if we use about 200 $\textbf{k}$ points in the IBZ and the
IEC converges faster than the total energy as a function of $\textbf{k}$
points in the IBZ. 

\section{Results and Discussion}
\subsection{Interlayer exchange coupling}
The total energies of the Fe/Nb multilayers have been calculated for para, ferro and 
antiferromagnetic ordering of Fe atoms for each Nb spacer layer thickness 
within the LSDA 
and the GGA exchange correlation potentials. It is found that the total energy for 
paramagnetic multilayers is always higher than the total energy of the other two 
magnetic configurations in both the LSDA and GGA potentials.

The interlayer exchange coupling of Fe/Nb multilayer 
is shown in Fig. 3 for the LSDA and GGA potentials, as a function of Nb spacer layer 
thickness. The IEC flips from ferromagnetic to antiferromagnetic at around 2.5 
\AA\ and 8.0 \AA\ and then to ferromagnetic coupling at around 4.0 \AA\ and 
10.5 \AA\ for the LSDA. The period of oscillation is found to be 6.0 
\AA\ which is in good agreement with the experimental value in the preasymptotic
region. In the experiment, the preasymptotic value is about 6.5 \AA\ while the asymptotic period is found to be 9.0 \AA\ \cite{kn:Mattson}.
The IEC has similar trends as in the ASW work \cite{kn:Sticht} except the period
of oscillation is slightly larger than the ASW work. We have taken an optimized
lattice constant which is smaller than the ASW work as well as we have 
considered only 1ML of Fe in our calculation while Sticht $\it{et}$ $\it{al.}$ \cite{kn:Sticht} 
have taken 2ML of Fe. This may be a possible reason for getting a different period 
from the ASW work \cite{kn:Sticht}. 

The IEC also shows the oscillations for the GGA potential but it does not change to
antiferromagnetic coupling and the period of oscillation is found to be the 
same i.e. 6.0 \AA\ . This indicates that the GGA favors ferromagnetic ordering 
in this system and it is in
good agreement with the experimental results \cite{kn:Mattson} where the IEC 
is found to be ferromagnetic for Nb thickness less than 14.0 \AA\ . Our 
calculations are restricted upto 11.0 \AA\ of Nb thickness. Similar behaviour 
has been seen in Fe/Au system where it was found that 2ML of Fe showed FM
coupling only \cite{kn:Yoshihara} while for higher number of Fe ML, FM as well 
as AF coupling \cite{kn:unguris} was observed.
This is consistent with the fact that in pure Fe, GGA gives the 
ferromagnetic ground state while the LSDA gives the paramagnetic ground state.
The magnitude of IEC is 4.45 mJ/m$^2$ for the LSDA and 22.64 mJ/m$^2$ for 
the GGA potential, which is very large compared to the experimental value of 0.034 
mJ/m$^2$. Similar trend has been reported earlier in
$\it{ab}$ $\it{initio}$ works \cite{kn:Mirbt,kn:Mirbt1,kn:unguris} on Fe/Cr, Fe/Au,
Co/Cu multilayer 
systems where the theoretical value of the IEC has been found to be 2 to 3 orders of 
magnitude higher than the experimental value. 
The GGA gives higher value of the IEC than the LSDA which is related to the 
fact that the GGA favors ferromagnetic ordering in this system. 
This is further
supported by our calculation of energy difference between ferromagnetic and 
antiferromagnetic states in bcc Fe using the GGA and LSDA. Using the GGA we get the 
energy difference of 0.49 eV while in the LSDA the energy difference is 0.39 eV.

The higher magnitude of the IEC as compared to
the experimental results
can be attributed to the following reasons: (1) The roughness at interface has
been neglected in our calculations. 
Bruno $\it{et}$ $\it{al.}$ \cite{kn:Bruno} studied the effect of interfacial
dislocations, interfacial roughness and lattice strain on the period and 
the amplitude of the IEC within the RKKY approach. They found that the interfacial dislocations 
reduce amplitude of the IEC. The reduction was more in (111) direction than
(001) direction in Cu/Co and Au/Co cases. The interface roughness causes 
fluctuations in the spacer layer thickness and also breaks in-plane translational
symmetry. As a result the IEC is strongly reduced and the period is increased.
In Fe/Nb multilayers, which have about $13\%$ lattice mismatch at the
interfaces, this effect is probably larger compared to 
the multilayers, which have matched interfaces such as Fe/Cr, Co/Cu, Fe/Au etc.
The interdiffusion and intermixing at the interface causes 
formation of FeNb alloy at the interface \cite{kn:Mattson}.
This will further contribute to roughness and suppress the IEC.
(2) The magnitude of the IEC in experimental 
study \cite{kn:Mattson} has been reported for higher thickness of Nb layer but 
our calculations are limited to 11.0 \AA\ Nb layer thickness.

\subsection{Magnetic properties} 

We find that the Fe layers induce the magnetic moments on Nb atoms. The 
induced moment on Nb atoms at interface layer is opposite to that of Fe atoms.
In Fig. 4, we show the induced magnetic moments per atom on Nb spacer layers 
using the GGA 
when Fe atoms are ferromagnetically ordered in $FeNb_{7}$ multilayers system. 
The induced magnetic moments show oscillations although 
the amplitude of the oscillations away from the interface is quite small.
The induced magnetic moments at the interface is 0.17 $\mu_B$ while at the 
middle layer it is 0.01 $\mu_{B}$ in $FeNb_{7}$ multilayer. We notice 
that the amplitude of the oscillations increases with the increase in number of 
Fe layers. This is similar to the nature of induced magnetic moment in the 
spacer layer seen in 
other systems \cite{kn:Stoeffler}. It is also seen that the Nb atoms have higher 
magnetic moments on the interfacial Nb spacer layer.
These induced magnetic moments are reduced in the LSDA 
calculations. The magnetic moment at the interface changes to 0.09 $\mu_B$ and 
zero magnetic moment at the middle layer of Nb in the LSDA. The explanation is given in 
next section for the difference of magnetic 
moments of Nb for both the LSDA and GGA functionals.

In Fig. 5, we plot the change in the magnetic moment per atom in Fe layers 
as a function of Nb spacer layer and find that it also shows oscillations in
the LSDA as well as GGA. Also shown in the inset are the experimental results
of Fe/Nb multilayers \cite{kn:Mattson}.
The magnetic moment of Fe is reduced at the interface as compared to the bulk 
value, in agreement with the experimental results. In case of the GGA potential, the magnetic moment of Fe varies from
1.32 $\mu_B$ to 0.90 $\mu_B$ while in LSDA, it 
varies from 1.12 $\mu_B$ to 0.46 $\mu_B$ depending on the Nb layers in the
system. 
In both cases the magnetic moment of Fe is  
smaller than the magnetic moment in bulk Fe (2.24 $\mu_B$). 
Our GGA calculations show that the magnetic moment of Fe at 
the interface decreases rapidly with Nb thickness till 5.0 \AA\ and is almost constant
between 5.0 \AA\ to 10.0 \AA\ and then shows a sharp increase around 
10.0 \AA\ . We note that the GGA results
show much better agreement with the experiment. In particular, the sharp rise of 
the magnetic moment around 10.0 \AA\ is well reproduced by the GGA results.
We also find the magnetic moment of Fe, in case of $Fe_{2}Nb_{2}$, varies
 from 2.16 $\mu_B$ 
to 2.00 $\mu_B$ and the induced magnetic moment on Nb layer also changes 
from 0.54 $\mu_B$ to 0.22 $\mu_B$. Thus the reduction in magnetic 
moment of Fe atoms at interface depends on Nb spacer thickness and magnetic 
moments of Nb spacer layer also depends on Fe layer thickness. 

\subsection{Electronic structure}

To understand the behaviour of the magnetic moments of Fe and Nb, we have examined 
the d-band density of states (d-DOS) of Fe and Nb in $Fe/Nb_{n}$ multilayer 
system. The ferromagnetic d-DOS for Fe in bulk and $Fe/Nb_{4}$ multilayers 
are shown in Fig. 6. This figure shows that the difference between up and down
electrons in bulk Fe is more compared to the corresponding difference in 
$Fe/Nb_{4}$ multilayers. Thus the magnetic moment of Fe in $Fe/Nb_{4}$ 
multilayers is smaller compared to the bulk value. Also we note from the figure 
that the down spin d-DOS in $Fe/Nb_{4}$ is shifted to lower energy compared to 
the corresponding d-DOS in bulk Fe. This shows that the exchange splitting in
$Fe/Nb_{4}$ is smaller than that in bulk Fe. The figure also shows a 
comparison between the LSDA and GGA d-DOS. We see from the figure that 
exchange splitting in the GGA is larger compared to the LSDA.

In Fig. 7, we have shown the d-DOS of Nb layers at the interface and at the 
middle layer for
ferromagnetic $Fe/Nb_{4}$ multilayers using the GGA. The bulk d-DOS of Nb is also 
shown. The layer at interface is denoted by Nb1 and the middle layer by Nb2.
The interface and middle layers have different d-DOS compared to the bulk d-DOS of Nb. 
The d-DOS at the interface 
layer has peaks between -3.0 eV 
to 2.0 eV while the middle layer has much smaller d-DOS in this energy range.
This shows some hybridization of d-states of Nb and Fe at the 
interface layer while this hybridization is almost absent in the middle layer.
This is 
consistent with the result in Fig. 4 which shows that the Nb interface layer 
near the Fe layer has higher magnetic moment as compared to the layers far
from Fe layers. Thus the induced magnetic moment on Nb layer is a consequence 
of the hybridization between Fe and Nb d-states. Nb magnetic moment shows 
oscillations as shown in Fig. 4  because Fe layer induces a spin polarization 
in Nb layers which oscillates due to Friedel oscillations \cite{kn:fridel}. This is consistent 
with the earlier result for Fe/Mo multilayers where magnetic moment on Mo also
shows Friedel oscillations \cite{kn:Mirbt1}. Mirbt $\it{et}$ $\it{al.}$ \cite{kn:Mirbt1} have also
found the reduction in the Fe magnetic moments which is same in our case.  
The mechanism of the oscillation of Fe magnetic moments is not clear and open 
for further work.

\subsection{Comparison with Models}
Several models \cite{kn:Bruno,kn:Mirbt1,kn:himpsel,kn:hari} have been presented to 
explain the period of oscillations of the IEC. The RKKY and QW models express
the period in terms of the nesting vectors at the Fermi surface. The RKKY and 
the QW models give the same period because of the dependence on the dimensions 
of the Fermi surface of the space material. We shall now discuss the results
of these models and compare them with our results.

\subsubsection{RKKY Model}
In the RKKY theory, one magnetic layer polarizes 
the spacer layer and the other magnetic layer interacts with this. 
Since the spin polarization is oscillatory as a function of spacer layer
thickness, the IEC also oscillates.
The period of oscillation is given by $2\pi/\textbf{k}$, where \textbf{k} is 
the spanning vector \cite{kn:Fs} across the Fermi surface of the spacer layer.
Therefore, to calculate the period of oscillation of the IEC we have calculated
the Fermi surface of Nb which has many critical 
spanning vectors corresponding to many oscillation periods. 
The calculated 
values of spanning vectors in [100] plane are 0.79, 0.63, 0.53, 0.46, 0.35, 0.20 and 0.15 in 
unit of $\pi/d$, where d is the distance between the lattice planes
along (001) direction. The corresponding 
oscillation periods for these spanning vectors are 4.20 \AA\ , 5.43 \AA\ , 
6.20 \AA\ , 7.24 \AA\ , 9.31 \AA\ , 16.29 \AA\ and 21.73 \AA\ which are in good 
agreement with earlier calculations \cite{kn:Fs}.
The oscillation period of one of these spanning vector is 6.20 \AA\ , which is 
in good agreement with our FLAPW calculation. 
Because our FLAPW calculations are restricted up to 7ML of Nb spacer layer 
thickness, we could find only one period of oscillation. 

\subsubsection{QW Model}
We have also calculated the period of oscillation of the IEC using the quantum well 
(QW) model \cite{kn:Fs} even though within first order perturbation theory the
RKKY and QW models are equivalent. In this model 
the Fermi energy oscillates as a function of the well 
thickness. The period of the oscillation is given by the wave vector at the 
Fermi energy in the well.  We consider the free electron model where the 
standing wave form in QW if $\lambda = 2L/n$, where $\lambda$ is the wavelength, 
$L$ is the thickness of the well and $n$ specifies the energy level. $L$ is defined as 
$2L = Nd$, where $N$ is the number of spacer ML and $d$ is the interlayer spacing. 
Thus using the above relations and $k = 2\pi/\lambda$, we get the condition 
$k = 2\pi/Nd$. For a free electron in the well, the energy can be written as 

\begin{equation}
E =\frac{\hbar^2 k^2}{2m} = \frac{\hbar^2}{2m}(\frac{2\pi n}{Nd})^2
\end{equation}

As shown in Ref. \cite{kn:Fs}, energy $E$ shows oscillation as a function of $N$. Since the density of states (DOS) for this 
case varies as $E^{1\over 2}$, it will also show oscillation as a function 
of N at Fermi level.
The period of oscillations can be predicted by 
analyzing the peak positions of QW states in the electronic density of states 
\cite{kn:Fs, kn:himpsel}. In Fig. 8, we have plotted 
the variation in s and p DOS in the Nb spacer layer at Fermi energy as a 
function of Nb spacer layer thickness, which shows a peak at 3.0 \AA\
 and at 9.0 \AA\ .
This gives 
period of oscillation to be 6.0 \AA\ which is in good agreement with the result 
obtained from our FLAPW calculation and the RKKY model.

Since both of these models gives the same results for the period of the IEC, 
the comparison of our results with the model results gives no clue regarding the
mechanism of IEC. This question has been addressed by Schilfgaarde
and Harrison \cite{kn:hari} and 
Mirbt $\it{et}$ $\it{al.}$ \cite{kn:Mirbt1} in connection with Fe/Cr/Fe and 
Fe/Mo/Fe layers. In the RKKY theory, the amplitude of the coupling is bilinear in the 
magnetic moments on two sides while in QW model the amplitude does not depend
on the magnetic moments. To analyze this, the calculated IEC was fitted to the 
asymptotic limit of the RKKY expression and the amplitudes were calculated.
For the fit, results for spacer layer thickness less than 5ML were ignored as the expression
used for the fit corresponded to the asymptotic limit. Since the electronic 
structures of Nb and Mo are similar \cite{kn:Mirbt1,kn:morzi,kn:jani}, Fe/Nb system may show a behaviour similar 
to Fe/Mo system. For the Fe/Mo 
system, Mirbt $\it{et}$ $\it{al.}$ \cite{kn:Mirbt1} fitted their results up to
20ML and found that the amplitude is proportional to the magnetic moments
squared of Fe for 3ML, 4ML and 10 ML caliper while the IEC has no dependence on
magnetic moments for 2ML caliper. Thus their analysis does not favor either
the RKKY or the QW model. In our case the maximum thickness is 7ML and therefore
this kind of fit to the asymptotic expression is not possible at present and
will be taken up in future work.
        
\section{Conclusions}

We have calculated the interlayer exchange coupling between Fe layers when separated by Nb 
spacer layers using the self-consistent FLAPW method with the LSDA and GGA exchange 
correlation potentials. We observe an oscillating 
exchange coupling as a function of Nb spacer layer thickness with a period of 
6.0 \AA\ for the LSDA as well as GGA in good agreement with the preasymptotic 
value observed in the experiment. It is close to earlier calculated 
period 4.5 \AA\ by the ASW method. 
Our calculated period is also in good agreement with the period calculated  
using the RKKY and QW models.
The LSDA calculations show an AF coupling 
at 2.5 \AA\ and 8.0 \AA\ thickness of Nb space layer but the GGA calculations 
favor ferromagnetic coupling for all Nb spacer layers. 
We find that the induced magnetic moments 
on Nb atoms show oscillatory behaviour. The Fe magnetic moment also shows strong 
oscillations as a function of Nb spacer layer thickness. The GGA gives higher 
magnetic moments for Fe and Nb compared to the LSDA. 
The GGA results show better agreement with the experiment than the LSDA results.
The reduction in magnetic moments of Fe and the
induced magnetic moment on Nb atoms is found to be a consequence of the hybridization 
between Fe and Nb d-bands. 

\begin{center}
\textbf{Acknowledgment}
\end{center}

We thank Profs. M. K. Harbola, S. Auluck, R. C. Budhani and A. K. Majumdar 
for helpful discussions. This work was supported 
by the Department of Science and Technology, New Delhi, via project no. SP/S2/M-51/96.

\newpage

$\bf{FIGURE}$ $\bf{CAPTIONS}$

Fig. 1. Supercell used for $FeNb_{4}$ multilayer.

Fig. 2. Interlayer exchange coupling, J(meV) for $FeNb_{n}$ multilayers as a function of 
$\textbf{k}$-points in the IBZ for different number of Nb layers for
(a) LSDA and (b) GGA. The calculated results are shown by filled squares, 
circles and stars.

Fig. 3. Interlayer exchange coupling, J (meV) for $FeNb_{n}$ multilayers as a function of Nb 
spacer layer thickness ( \AA\ ) for the LSDA and GGA. 

Fig. 4. Induced magnetic moment on Nb spacer layer in ferromagnetic $FeNb_{7}$ multilayer.
 The Fe atoms are placed at positions 0 and 8. The filled squares show the calculated results.

Fig. 5. Magnetic moments on Fe atoms in $FeNb_{n}$ multilayers as a function of 
Nb spacer layers. The dots correspond to the GGA and filled squares to the LSDA results. The experimental results of Mattson $\it{et}$ $\it{al.}$ \cite{kn:Mattson} are shown in the inset.

Fig. 6. The top panel shows the partial density of states of spin-up and spin-down d-bands for ferromagnetic bulk Fe using the GGA. The bottom panel shows 
the partial density of states of spin-up and spin-down d-bands for Fe in 
ferromagnetic $Fe/Nb_{4}$ system using the LSDA and GGA.

Fig. 7. The top panel shows the partial density of states of spin-up and 
spin-down d-bands for Nb in ferromagnetic $Fe/Nb_{4}$ system. Nb1 denotes the 
d-DOS at the interface while Nb2 denotes the d-DOS at the middle layer. The bottom 
panel shows the d-DOS for bulk Nb using the GGA.

Fig. 8. Density of states of s and p states of Nb at Fermi level as a function of Nb spacer layer thickness.

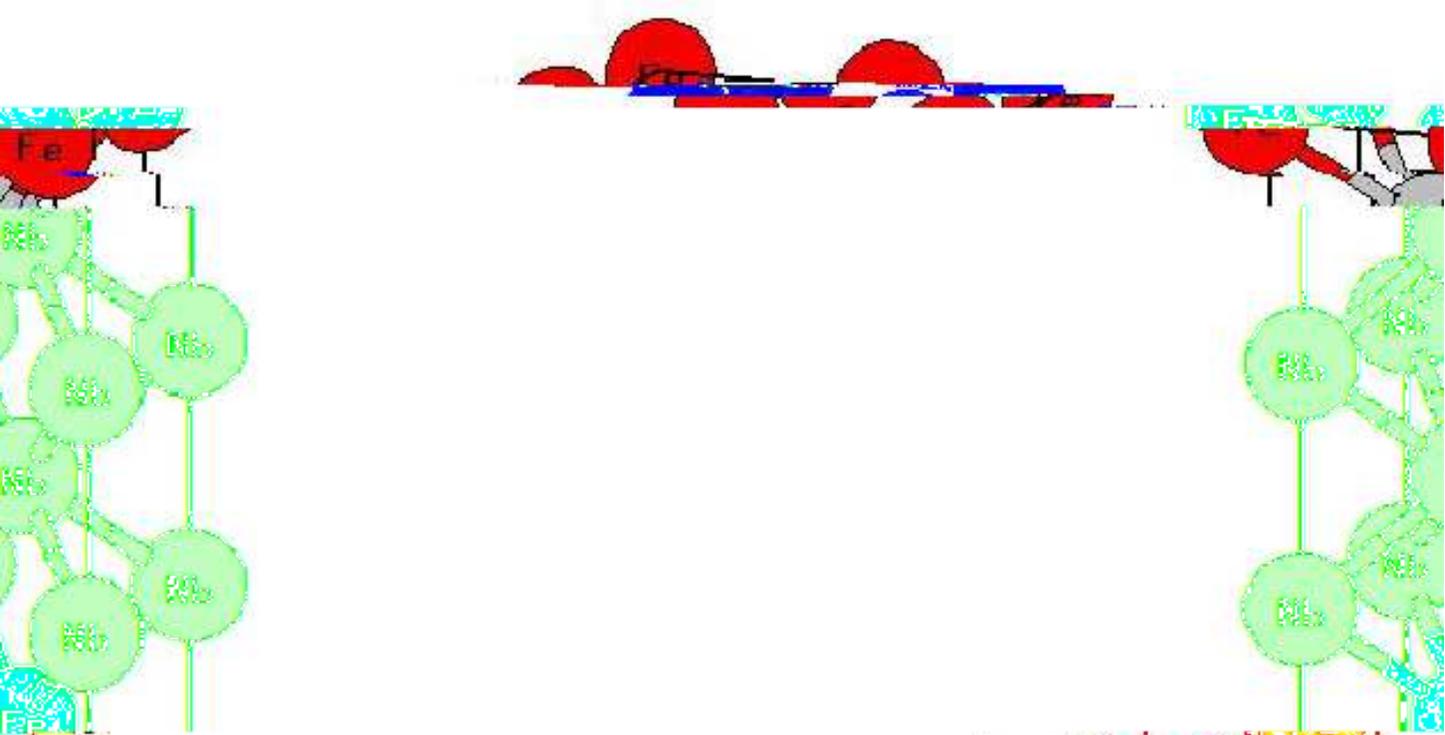
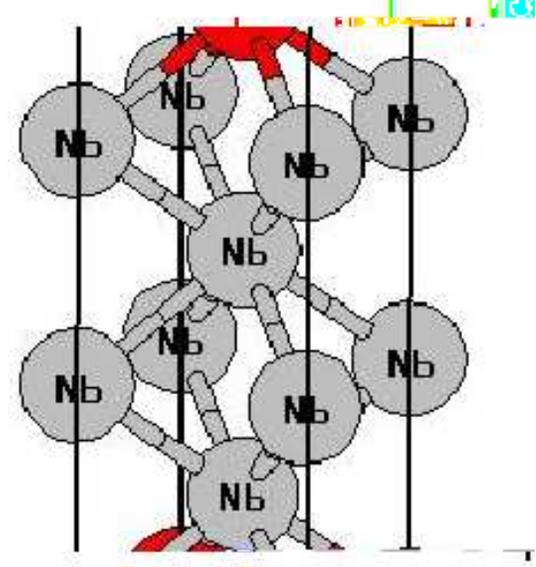
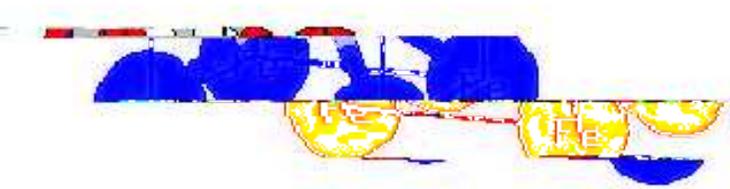
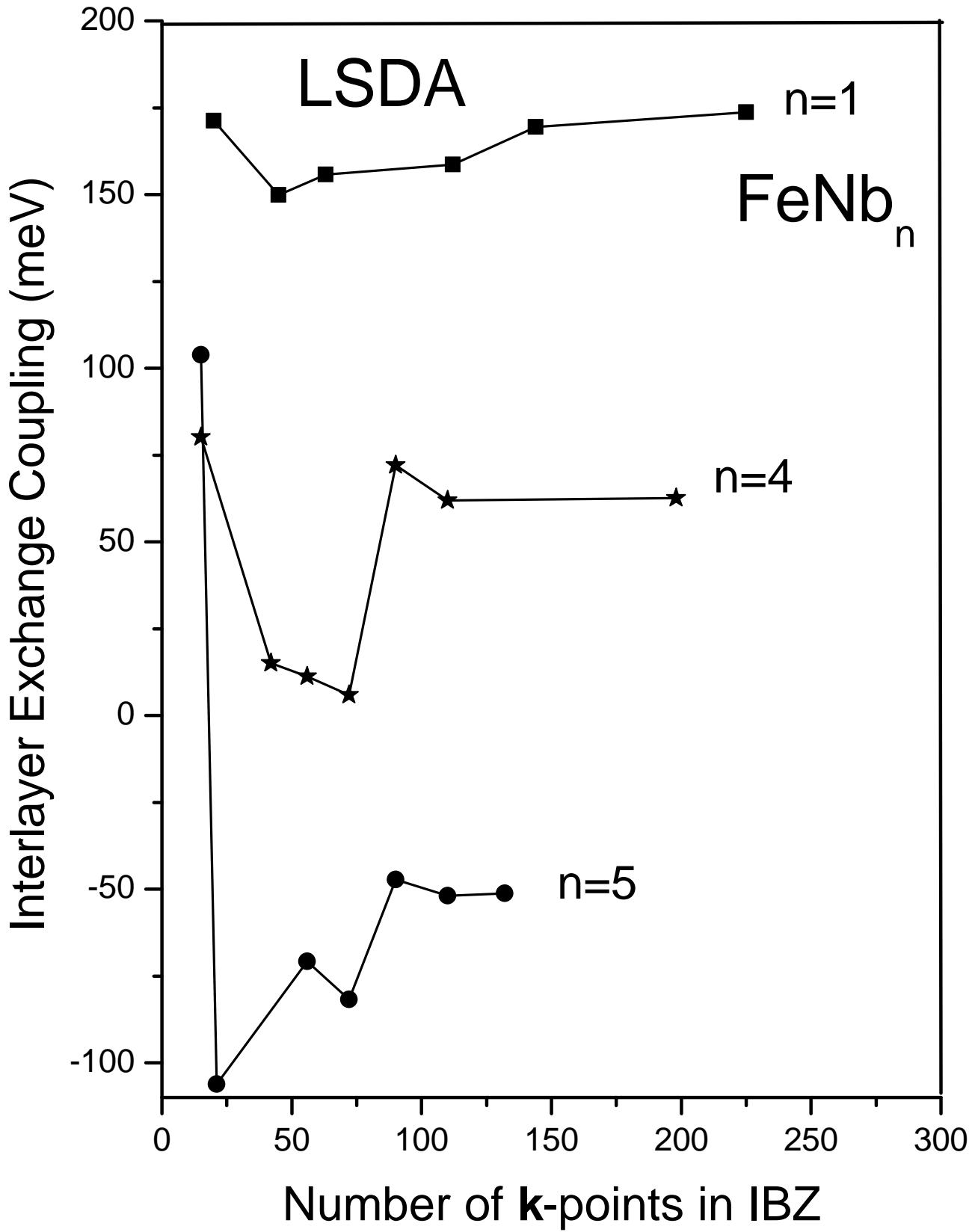
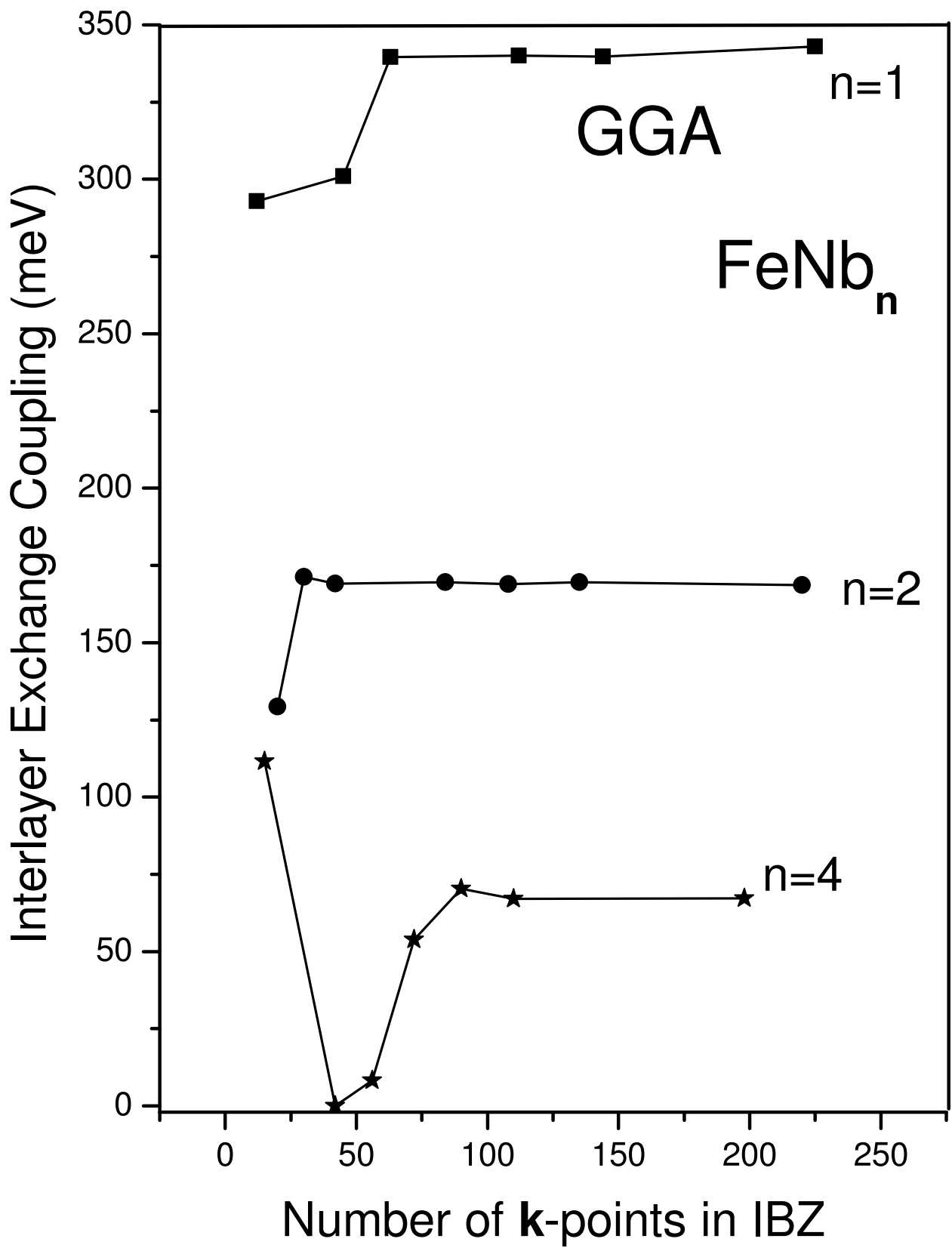
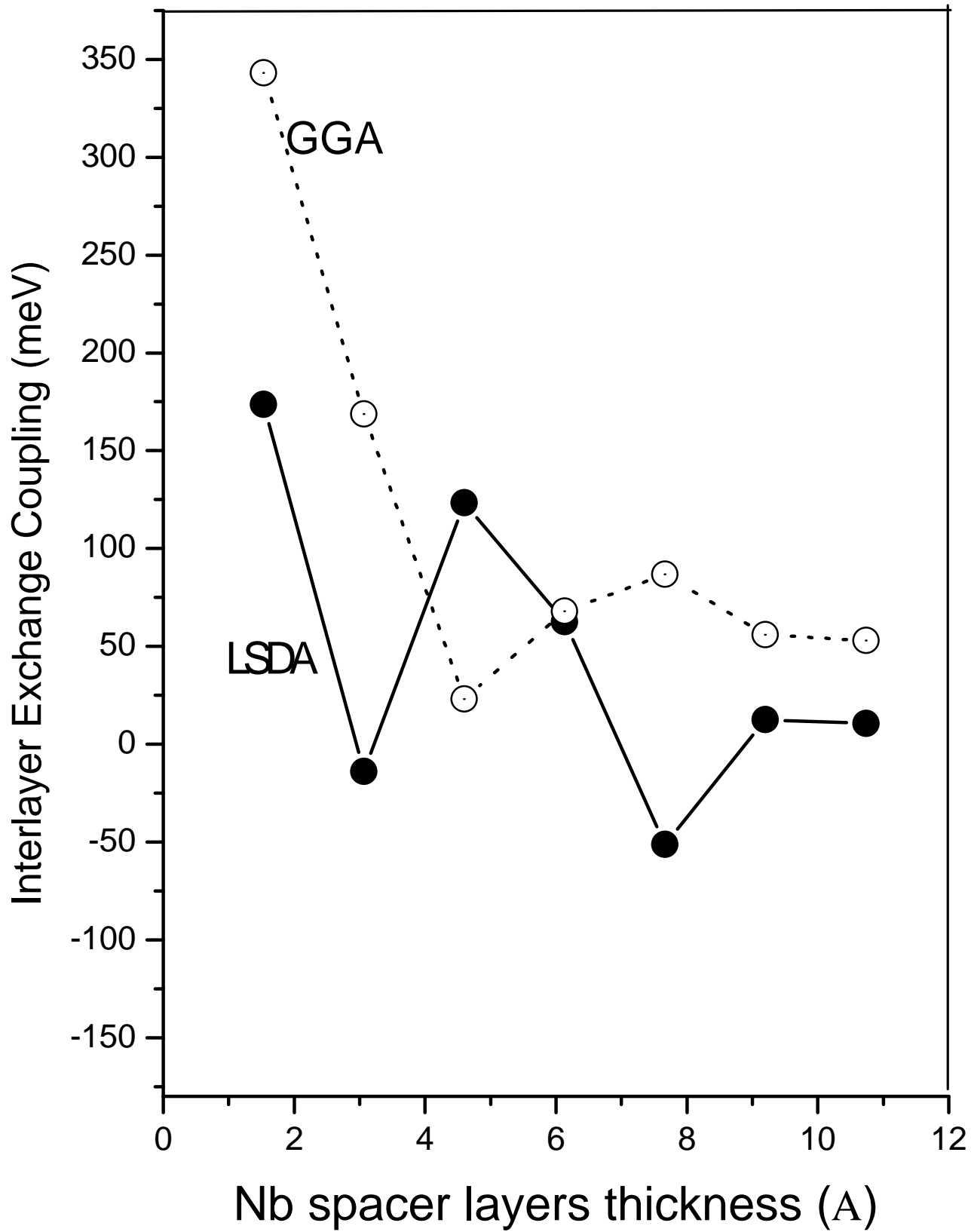
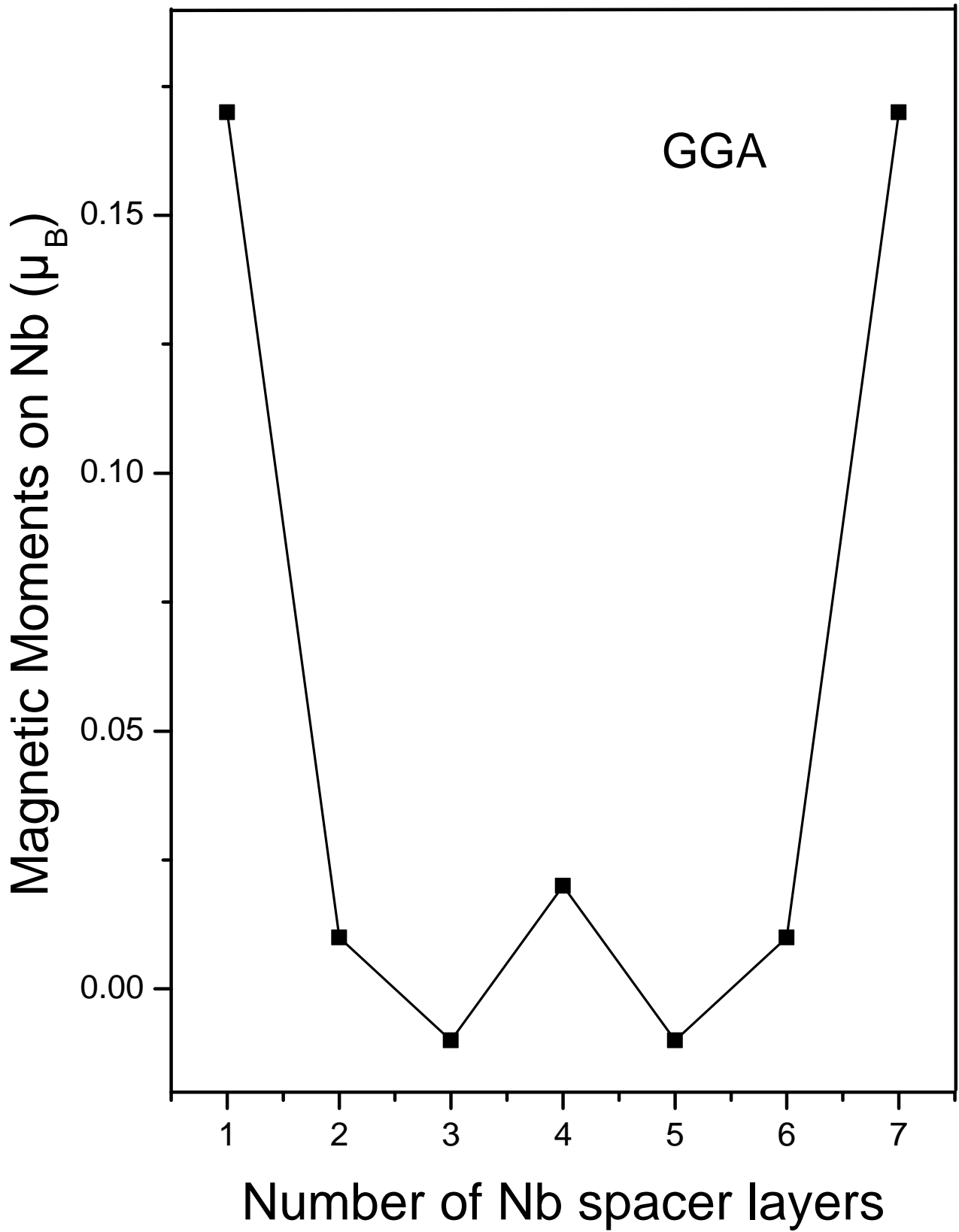
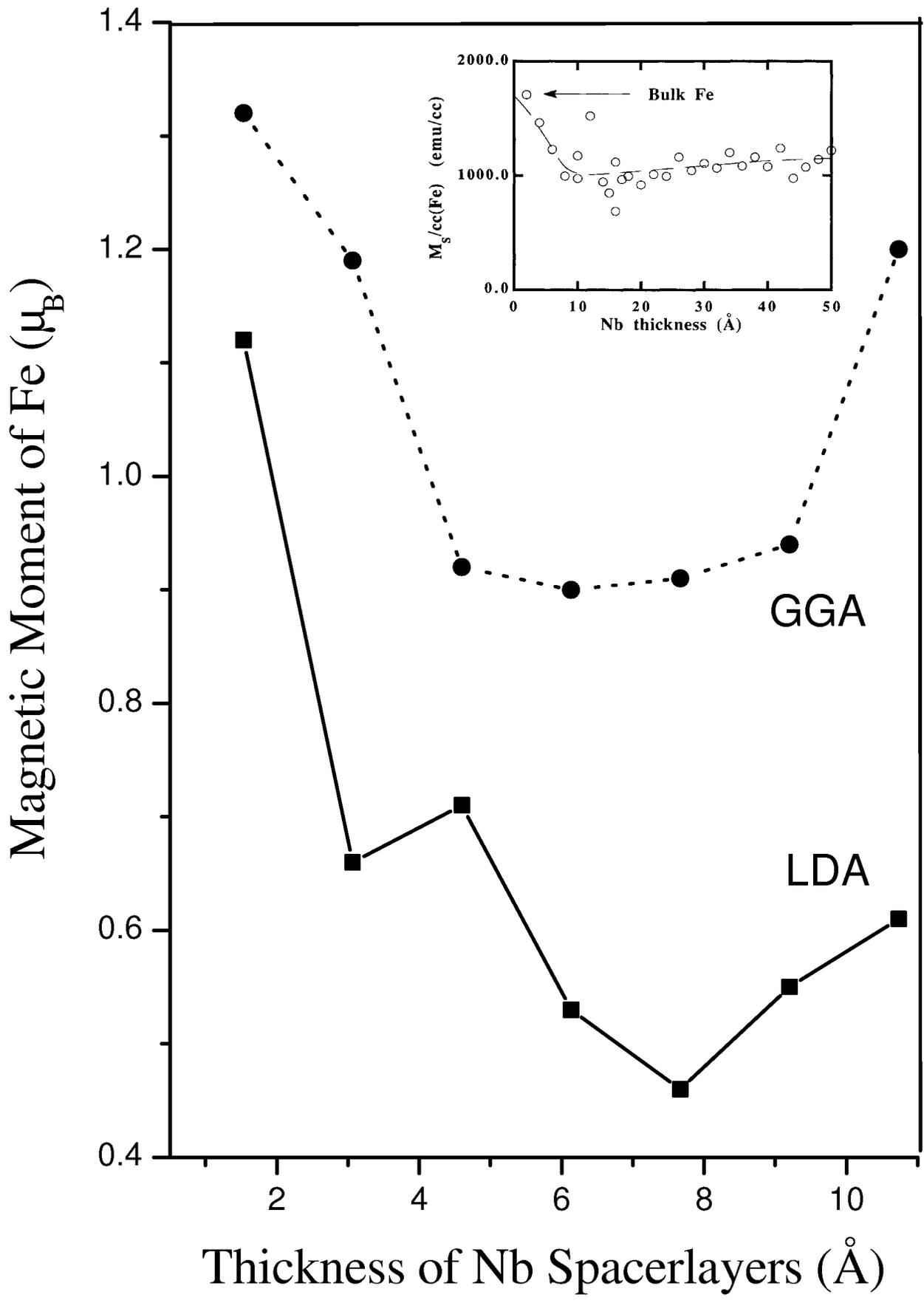
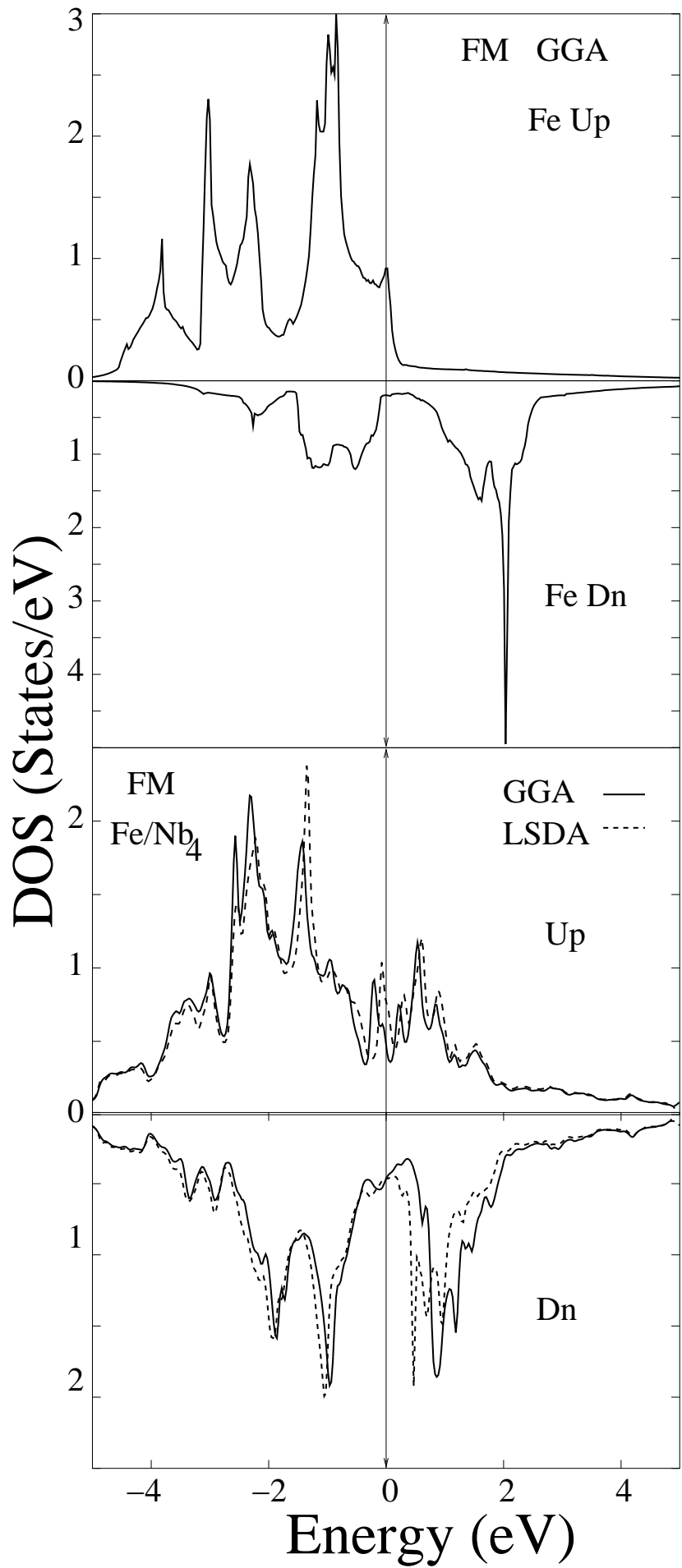
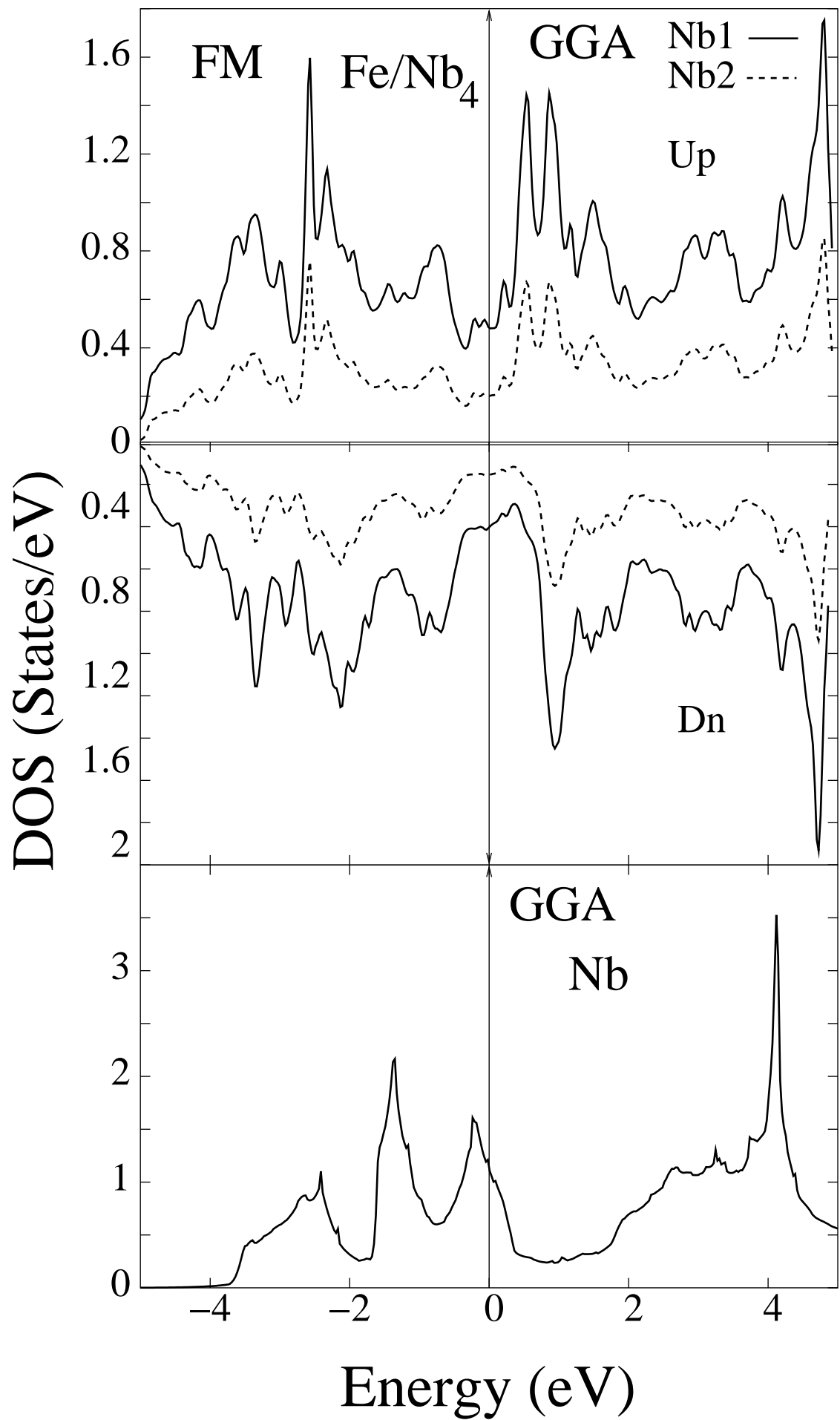
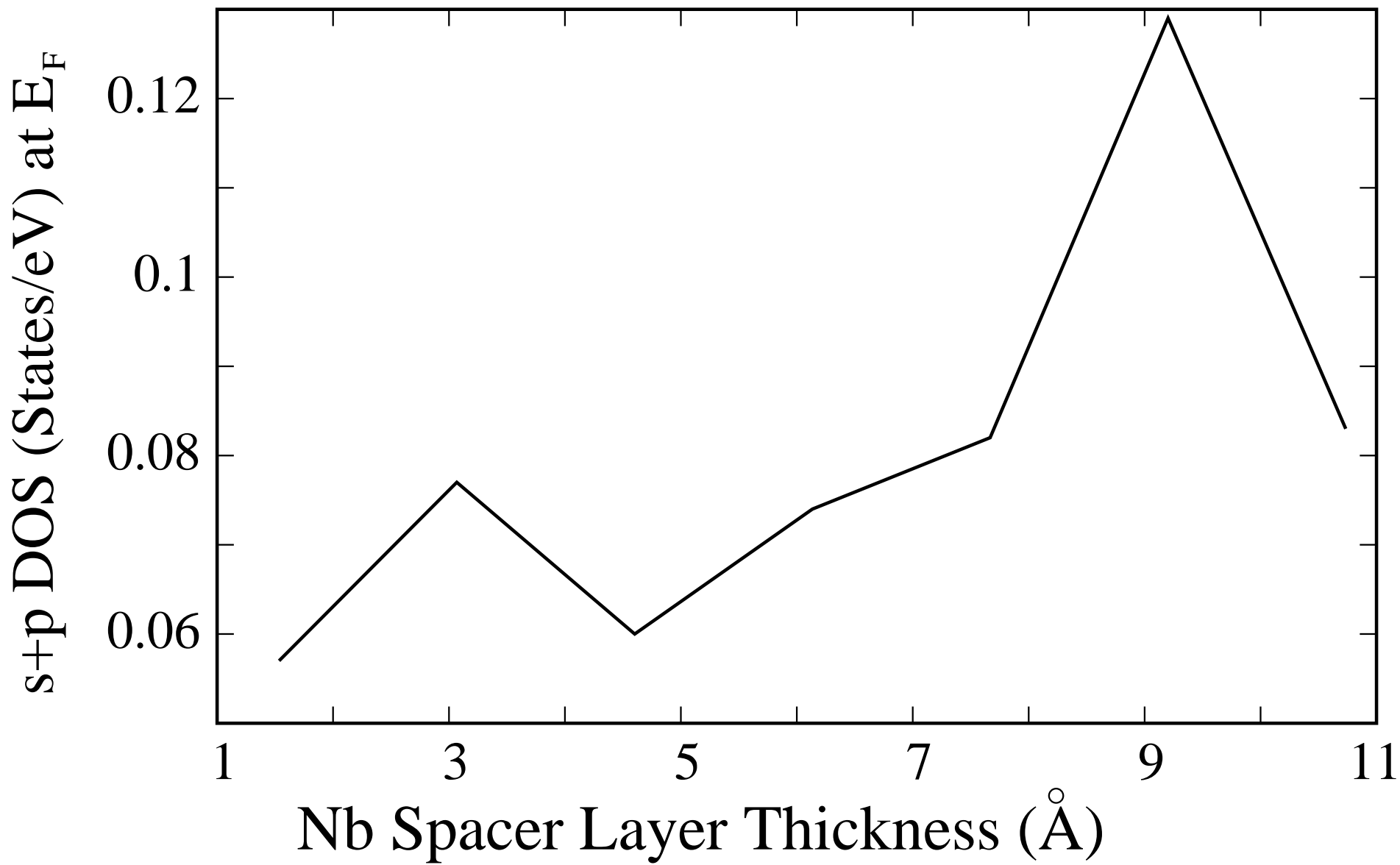
\end{document}